\documentclass[a4paper]{spie}  
\usepackage[]{graphicx}
\usepackage{bm}

\title{Quantum spin Hall effect in 2D topological insulators} 

\author{E. B. Sonin
\skiplinehalf
Racah Institute of Physics, Hebrew University of
Jerusalem, Jerusalem 91904, Israel}

\authorinfo{Further author information: E-mail: sonin@cc.huji.ac.il}

\begin{document} 
  \maketitle 


\newcommand{\be}{\begin{equation}}
\newcommand{\ee}[1]{\label{#1}\end{equation}}
\newcommand{\bem}{\begin{eqnarray}}
\newcommand{\eem}[1]{\label{#1}\end{eqnarray}}
\newcommand{\eq}[1]{Eq.~(\ref{#1})}
\newcommand{\Eq}[1]{Equation~(\ref{#1})}

\begin{abstract}
 The original motivation of great interest to topological insulators was the hope to observe the quantum spin Hall effect.  Therefore if a material is in the topological insulator state they frequently call it the quantum spin Hall state. However, despite impressive experimental results confirming the existence of the quantum spin Hall {\em state}, the quantum spin Hall {\em effect} has not yet been detected.  After a short overview of what was originally suggested as the quantum spin Hall effect (quantum spin conductance determined by the topological Chern number) the paper analyzes the crucial role of the boundary condition on the observation of the effect and finally discusses whether and how the quantum spin Hall effect could be observed.
\end{abstract}



\section{Introduction}
\begin{figure}
\begin{center}
\begin{tabular}{c}
   \includegraphics[height=7cm]{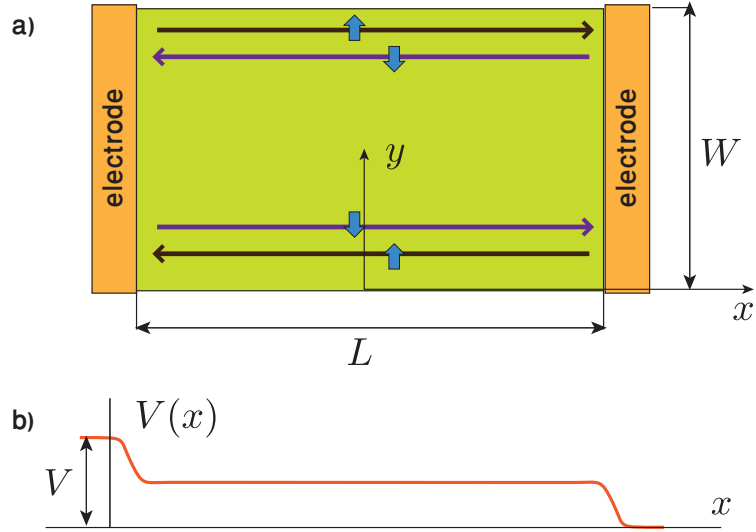} 
\end{tabular}
\end{center}
 \caption{ Voltage biased topological insulator. (a)Edge states. Wide arrows show spin direction (spin quantization axis is not in the plane as in the figure). At the upper edge  rightmovers have spin s up while leftmovers have spins down.  At the lower edge directions of motion are opposite. (b) Voltage variation between electrodes. The voltage varies only near electrodes whereas inside the sample there is no electric field because of ballistic transport along edge channels.  }
 \label{figTI}
 \end{figure}

Topological insulators have recently attracted great attention and their study is developing  into a new exciting area of condensed matter physics\cite{KM,KMa,BHZ,Koenig,KoenigE,Joel,Hasan,Hsi,Xia,Chen}.  The key feature of topological insulators  is the presence of helical edge states in 2D systems (or surface states in 3D systems). The edge states cross the whole forbidden gap  separating the valence-band and the conduction-band bulk continua. Helicity of edge states means that electrons with the same spin can move only in one direction, which is opposite for two  spin directions.  As a result,  the edge states are robust against elastic backscattering, which conserves spin, and the electron transport along edges becomes ballistic.  The edge state of the 2D topological insulator were experimentally detected in the HgTe quantum well \cite{KoenigE} by studying charge transport. It was demonstrated that at the quantum well thickness exceeding the critical value 6.3 nm there was an interval of gate voltages where the conductance reaches the quantum conductance value independently of the sample width $W$ (see Fig.  \ref{figTI}). This is a clear evidence of the ballistic transport through edge states while the main bulk is not conducting. The topological insulators states were also detected in 3D materials  \cite{Hsi,Xia,Chen}. Though nowadays topological phases are mostly investigated in solids they were already well known in superfluid $^3$He from 80s \cite{Vol}.

Originally 2D topological insulators were introduced as systems, in which the quantum spin Hall effect (QSH) was expected  \cite{KM,BHZ}, and the state of the topological insulator is frequently called the QSH state. However, it  is necessary to clearly specify the definition of the spin Hall effect in general and the  quantum spin Hall effect in particular. Historically the phenomenon, which is called nowadays  spin Hall (SH) effect, was defined as an edge spin accumulation resulting from a bulk spin current transverse to an external electric field \cite{DP}. At the sample boundary, which fully reflects  all electrons, the spin current induced by the electric field must be compensated by a spin diffusion current. Spin diffusion is a dissipative process, which is related with non-equilibrium spin accumulation  at the edge (or spin orientation in the terminology of Dyakonov and Perel \cite{DP}).  Since spin accumulation is more accessible  for observation  than spin current, usually observation of spin accumulation is considered as an experimental confirmation of the SH effect \cite{Kato,Wund}. But later it was revealed that spin accumulation is not an ideal probe of bulk spin currents since spin accumulation is possible without bulk spin current, and, vice versa, spin current not necessarily results in spin accumulation  (see the review \cite{Adv} and references therein). This is a consequence of non-conserved total spin. In particular, at the equilibrium bulk spin currents cannot lead to spin accumulation, since the latter is accompanied by dissipation. Instead spin is transformed to an orbital moment at the sample edge (edge torque\cite{Mech}). The same is true for bulk spin currents in insulators: since an external electric field does not pump energy into the sample (no Joule heating) dissipative spin accumulation cannot be supported.

The adjective "quantum" was added to the SH effect in the topological insulators  in order to stress that the spin conductance (ratio of the bulk transverse spin current  to the electric field) was  universally determined by the topological Chern number for the 2D Brillouin zone \cite{BHZ}. Such a QSH effect has not yet been detected, since the author is not aware  of any measurement, which probes anything connected with the bulk spin conductance and no experiment up to now is able to detect the predicted quantum spin conductance in the topological phase.  Nevertheless, it is believed that the  QSH effect has already been observed \cite{Koenig}.  The 
ingenious experiments of K\"onig {\em et al.} \cite{KoenigE} definitely confirmed existence of ballistic edge states 
and revealed the quantum {\em charge}  conductance but there were no direct or indirect evidences of  quantum {\em spin}  conductance simply because nothing dependent on spin was measured. Though both, the ballistic edge  states (and the quantum charge conductance as a result of it) and the quantum spin conductance, originated from topology, the common origin does not make two phenomena identical. An essential progress in studying spin-related phenomena in topological insulators was recent detection of spin polarization of the edge states \cite{SpPol} using the inverse SH effect, but this new impressive achievement still does not provide any information on bulk quantum spin conductance, since the latter does not affect any measurement result. 
So the QSH effect, at least in its original  meaning, still waits its experimental confirmation. 

The present  paper elaborates this point of view. It 
starts from a short overview of the model by Bernevig, Hughes, and Zhang (BHZ) for the topological 
 insulator (Sec.~\ref{Mod}) and the definition of spin and spin current in this model  (Sec.~\ref{SSC}). In Sec.~ \ref{QSC} the quantum spin conductance for the periodic boundary  is discussed, which is determined by the topological Chern number and was originally  considered as the determining feature of the QSH effect. Section~\ref{RBC} shows that in the strip geometry with fully reflective lateral edges it is impossible to realize  the quantum spin conductance. But in this geometry there is non-quantum spin conductance, which is possible both in the topological and conventional-insulator phases of the BHZ model. Section~\ref{spAc} addresses spin accumulation (polarization) at edge states and argues that observation of spin polarization cannot be considered as demonstration of the QSH effect. The concluding Sec.~\ref{concl} discusses a possibility to detect the genuine QSH effect using the Corbino geometry and summarizes the present analysis: the QSH effect is still waiting its experimental confirmation.

\section{The topological-insulator model of Bernevig, Hughes, and Zhang}\label{Mod}

The simplest and therefore the most transparent model for studying fundamental features of the topological insulator is  the model suggested by Bernevig, Hughes, and Zhang (BHZ) for the topological 
 insulator in the HgTe quantum well \cite{BHZ,Koenig,zhou}. The model is a simplified version of the Kane model and its 4$\times$4 Hamiltonian is given by 
\begin{equation}
{\cal H}=\left(\begin{array}{cc} \hat H({\bm k}) & 0 \\ 0 &  \hat H(-{\bm k})^*\end{array} \right),
   \label{ham4} \end{equation}
where 
\begin{equation}
\hat H({\bm k})=\varepsilon(k)\hat I + d_i \hat \tau _i
       \label{ham2}\end{equation}
is a 2$\times$2 Hamiltonian, $\hat \tau_i$ are Pauli matrices of the pseudospin, $i=x,y,z$,  and $\hat I$ is a unit $2\times2$ matrix. Assuming that all essential processes occur at low $k$ close to the Brillouin zone center, the components $d_i$ of the vector $\bm d$  are
\begin{equation}
d_x=Ak_x,~~d_y=A k_y,~~d_z=\epsilon_0(k)=M-B k^2.
\end{equation}
Two components of the pseudospin in  any $2\times 2$  block of the Hamiltonian (\ref{ham4}) correspond to the valence (pseudospin up) and the conduction (pseudospin down) bands, which overlap in the  topological insulator phase at $M>0$. The off-diagonal linear in ${\bm k}$ terms in any block  lead to mixing of two original bands and to forming new bands separated by a forbidden gap. The conventional-insulator phase without edge states corresponds to the condition $M<0$.  Further we shall neglect $\varepsilon(k)$ in the Hamiltonian (\ref{ham2}) as not important for the outcome of the analysis \cite{Koenig}.

\begin{figure}
  \begin{center}
   \begin{tabular}{c}
   \includegraphics[height=11.5cm]{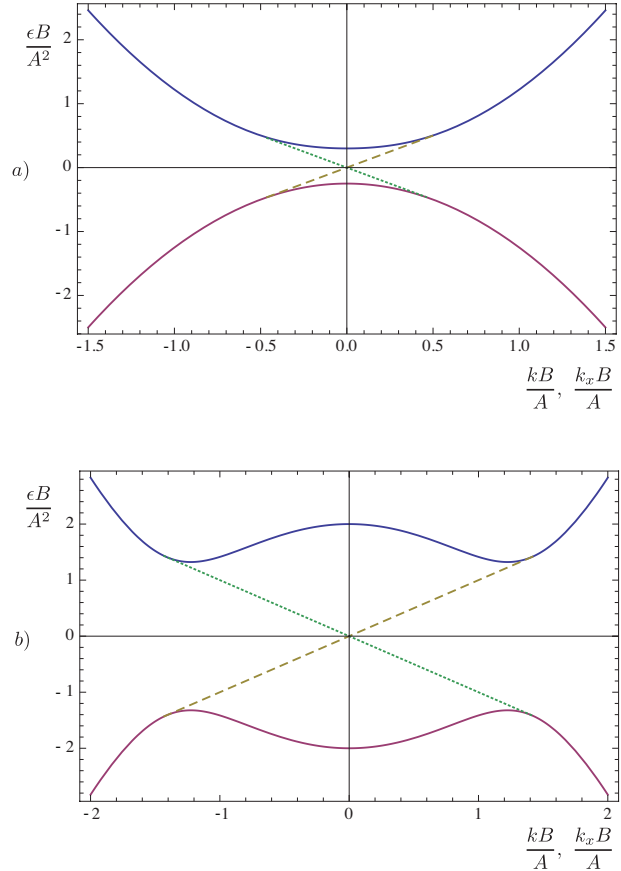}%
      \end{tabular}
   \end{center}
 \caption{ Band energy as a function of $k$ (solid lines) and edge state energy as a function of $k_x$ (dotted and dashed lines) in  a topological insulator ($M>0$).  Dashed and dotted lines correspond to opposite spin  directions [upper and and lower block of the Hamiltonian (\ref{ham4})].  a) $M=0.3 A^2/B<A^2/2B$; b) $M=2  A^2/B>A^2/2B$.}
\label{f1} \end{figure}

The lower block yields the states obtained from those for the upper block by the time-reversal transformation. Because of the absence of off-diagonal blocks in  the Hamiltonian (\ref{ham4}), one can analyze states for any block separately. We consider the upper block. 
The eigenstates in the ${\bm k}$ space  for the Hamiltonian (\ref{ham2}) are spinors
\begin{eqnarray}
{\bm \Psi}_\pm(k_x,k_y)={1\over \sqrt{2\epsilon}}\left(\begin{array}{c} \sqrt{\epsilon\pm  \epsilon_0}  \\\pm {A(k_x+ik_y) \over \sqrt{\epsilon\pm \epsilon_0}}  \end{array} \right),
\label{spinor} \end{eqnarray}
where $\epsilon=|{\bm d}|=\sqrt{\epsilon_0^2+A^2k^2}$, and the spinors ${\bm \Psi}_+$ and ${\bm \Psi}_-$ correspond to the energies $+\epsilon$ and $-\epsilon$ respectively. 
 At $M<A^2/2B$ the energy of the upper band  has a minimum at $k=0$ (Fig.~\ref{f1}a). At $M>A^2/2B$ the energy has a maximum at $k=0$,  whereas  the  minimum band energy $\epsilon_m = A\sqrt{M/B-A^2/4B^2}$ corresponds to $k_m =\sqrt{M/B-A^2/2 B^2}$  (Fig.~\ref{f1}b).

An edge state near the edge $y=0$ should be a a superposition of  two states of the same energy and  $k_x$: 
\begin{eqnarray}
{\bm \Psi}=\left[a_1{\bm \Psi}_+(k_x,k_{y1})e^{ik_{y1}y}+a_2{\bm \Psi}_+(k_x,k_{y2})e^{ik_{y2}y}\right]e^{ik_xx},
\nonumber \\
\end{eqnarray}
where $k_{y1}$ and $k_{y2}$ are two complex solutions  with positive imaginary parts of the bi-quadratic equation for $k_y$ following from the energy spectrum at fixed $\epsilon$ and $k_x$.

The condition ${\bm \Psi}=0$ at the edge $y=0$ yields two equations for $a_1$ and $a_2$, which have a solution if
\begin{eqnarray}
\frac{A(k_x+ik_{y1})}{\epsilon_e +\epsilon_0(k_1)}=\frac{A(k_x+ik_{y2})}{\epsilon_e+ \epsilon_0(k_2)}.
\label{DR}\end{eqnarray}
For $M>0$ this  equation is exactly satisfied if $\epsilon_e=-Ak_x$, which corresponds to the edge  state 
\begin{eqnarray}
{\bm \Psi} \propto \left(\begin{array}{c} 1 \\ 1 \end{array} \right) e^{ik_x x}\left(e^{-p_+y}-e^{-p_-y}\right),
 \label{edg}  \end{eqnarray}
where $p_\pm =A/2B \pm \sqrt{k_x^2-M/B+A^2/ 4B^2}$. Considering the lower block of the Hamiltonian (\ref{ham4}) one receives the spectrum $\epsilon_e=Ak_x$.

\section{Spin and spin current in the BHZ model} \label{SSC}

Starting the analysis of spin currents and accumulation, it is necessary to understand what  ``spin''  we would like to focus on. In HgTe the conduction band originates \cite{BHZ,Koenig} from a  $s$-type ($l=0$) atomic  orbital, and its total moment coincides with spin, whereas the valence band is related to a $p$-type ($l=1$) atomic  orbital and has the total moment $ j=3/2$ with its projection on the quantization axis (the axis $z$ normal to the insulator plane) $m_j=1/2$. This projection of the mechanical total angular momentum  was also called  spin in the previous works  \cite{BHZ,Koenig}. We shall adopt this extension of the term ``spin'' remembering that in general this is an effective spin   but not  genuine one. According to this definition the spins in the valence and the conduction bands in HgTe are equal to the electron spin $\hbar/2$. However, the mechanical moment would be relevant only if the theory addressed a mechanic effect, like that considered in Ref.~\cite{Mech}. If the goal is to describe electromagnetic phenomena like the Kerr effect or any other electrodynamic effect, one should use the magnetic moment, which depends on the Lande factor of the atomic orbital. The magnetic moments in the  the conduction and valence bands of HgTe are $\mu_B$ and $2\mu_B/3$, where $\mu_B=e\hbar/2mc$ is the Bohr magneton. 
In order to take it  into account we further generalize the definition of effective spin, introducing ``spins'' $s_c$ and $s_v$ for the conduction and the valence bands. If the analysis addresses a mechanical effect then $s_c=s_v=\hbar/2$ in HgTe quantum wells as was supposed earlier  \cite{BHZ,Koenig}. But if the magnetic moment is relevant  for the theory, we define effective spins  of the conduction and the valence bands as $ s_c=\hbar/2$ and $s_v=\hbar/3$, i.e., as magnetic moments of the bands divided by the electron gyromagnetic ratio. So in general the band effective spins $s_c$ and $s_v$ differ and the operator  of the effective spin in the BHS model is given by \cite{QSH}
\begin{eqnarray}
\hat s^z =\bar s^z \hat I +\Delta s^z \hat \tau_z,
   \end{eqnarray}
where $\bar s^z=(s_c+s_v)/2$, and $\Delta s^z = (s_v-s_c)/2$.  

Let us consider now the balance equation for  spin (the continuity equation with the torque term)  for an arbitrary quantum state in the BHS model.
The balance equation  can be derived from the Schr\"odinger equation as was explained in details for the Rashba Hamiltonian \cite{Adv}. Restricting ourselves with the $z$ component of the spin density $S_z={\bm \Psi}^\dagger  \hat s^z  {\bm \Psi}$, the balance equation is 
\begin{eqnarray}
{\partial S_z\over \partial t} +\nabla_\alpha J^z_\alpha=G^z,
   \end{eqnarray}
where the torque is 
\begin{eqnarray}
G^z=i \Delta s^z A \left\{ {\bm \Psi}^\dagger \cdot  [\vec \nabla\times {\bm \tau}]_z {\bm \Psi}
+  [\vec \nabla\times {\bm \tau}] _z {\bm \Psi}^\dagger \cdot  {\bm \Psi}\right\},
   \end{eqnarray}
and the spin current is given by
\begin{eqnarray}
 J_i^z={1\over 2} {\bm \Psi}^\dagger  \left\{ \hat s^z \hat v_i +\hat v_i \hat s^z \right\} {\bm \Psi} = \bar s^z {\bm \Psi}^\dagger  \hat  v_i {\bm \Psi} + \Delta s^z v_{0i}{\bm \Psi}^\dagger {\bm \Psi}.
\nonumber \\
\label{cur1}   \end{eqnarray}
Here 
\begin{equation}
\hat v_i={\partial \hat H({\bm k})\over \hbar \partial k_i}=v_{0i}(k)\hat \tau_z + A \hat  \tau _i
       \label{cur}\end{equation}
is the  group velocity operator and  $v_{0i}(k)=\partial \epsilon_0/\hbar\partial k_i $. 

The first term in the spin current \eq{cur1} differs from the charge current 
\begin{eqnarray}
 J_i= e {\bm \Psi}^\dagger  \hat  v_i {\bm \Psi} 
\nonumber \\
\label{ChCur}   \end{eqnarray}
by the constant factor $\bar s^z/e$. Only this term was taken into account in previous publications assuming  $ \Delta s^z=0$. In this approximation our effective spin is a conserved quantity and the torque $G^z$ vanishes. 
But in general the second term proportional to the difference of spins in two bands should not be ignored and leads to important effects as shown below.

An interesting consequence of helical edges  in the topological phase $M>0$ is a persistent spin current flowing around the sample. According to Eq.~(\ref{edg}), two  edge states  transport the average spin $\pm \bar s^z$,
and the spin current along the edge is:
\begin{eqnarray}
j^z = \bar s^z(n_\rightarrow+ n_\leftarrow) v_e.
 \label{edCur}  \end{eqnarray}
 where $v_e=d \epsilon_e/\hbar d k_x=A/\hbar$ is the group velocity at edge states. This current exists even in the equilibrium \cite{Butt}, when there is no external electric field and  the 1D densities $n_\rightarrow$ and $n_\leftarrow$ of right-moving and left-moving charge  carriers are equal. So this is another example of an equilibrium spin  current \cite{Adv}.

\section{The effect of an electric field: quantum spin conductivity} \label{QSC}
 
\begin{figure}
\begin{center}
\begin{tabular}{c}
   \includegraphics[height=7cm]{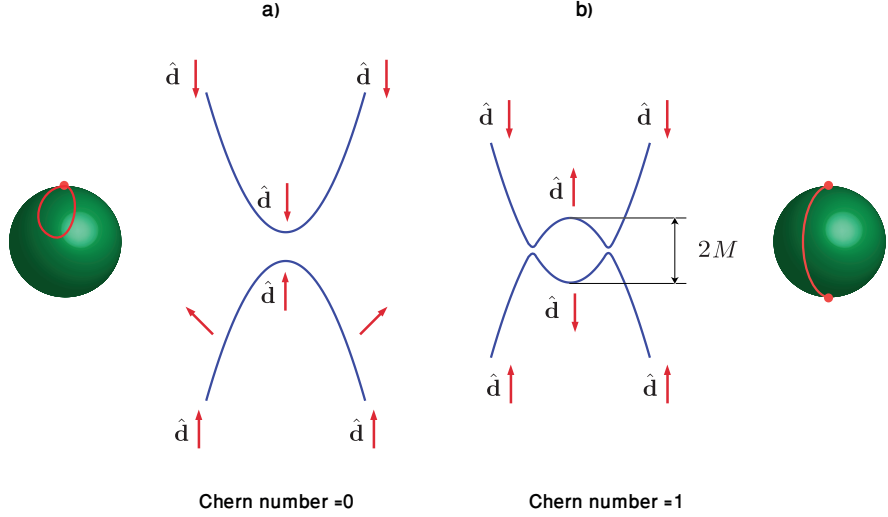} 
\end{tabular}
\end{center}
 \caption{ Mapping of the Brillouin zone on the spherical surface of the unit vector $\hat{ \bm d}=\bm d / d$. (a) The conventional insulator, $M<0$. Any path in the Brillouin zone starting in its center and ending at its border maps on the path in the $\hat{ \bm d}$ space starting and ending at the same pole of the sphere. The total area subtended 
by the vector $ \hat{\bm d}$ vanishes. 
 (b) The topological  insulator, $M>0$. Any path in the Brillouin zone starting in its center and ending at its border maps on the path in the $\hat{ \bm d}$ space starting at one pole and ending at the other. The vector $ \hat{\bm d}$ subtends the whole spherical surface with area $4\pi$. 
 }
 \label{figM}
 \end{figure}

The QSH effect follows from the calculation of the spin current induced by an electric field using the Kubo approach \cite{BHZ}. The approach is based on the perturbation theory. In the presence of the electric field the perturbation-theory expression for the quantum-state spinors are 
\begin{eqnarray}
\tilde{\bm \Psi}_\pm(k_x,k_y)= e^{ik_x x+ik_y y}\left\{{\bm \Psi}_\pm
+{i \hbar eE  \over 4 \epsilon^2}\hat v_x{\bm \Psi}_\pm\right\}.
\label{psiE}   \end{eqnarray}
The transverse spin current in this state follows from its definition by  \eq{cur1} and is equal to
\begin{eqnarray}
J_y^z(k_x,k_y) = \bar s^z\left(v_y +{i \hbar eE  \over 4 \epsilon^2}{\bm \Psi}_\pm^\dagger[\hat v_y,\hat v_x]{\bm \Psi}_\pm\right)
+  \Delta s^z v_{0y}
=\bar s^z\left(v_y +{ eE  \over 2 \hbar}{\cal G}\right)+  \Delta s^z v_{0y},
 \label{trCaur}  \end{eqnarray}
where the term 
\begin{eqnarray}
{\cal G}={A^2\over \epsilon^3}(\epsilon _0-\hbar k v_{0})= \hat{\bm d}\cdot \left[  {\partial \hat{\bm d}\over \partial k_x} \times  {\partial \hat{\bm d}\over \partial k_y}  \right]
      \end{eqnarray}
is responsible for the topological contribution to the current. Its integral 
\begin{eqnarray}
\int {\cal G} d{\bm k}=2\pi\left(1+{M\over |M|}\right)=4\pi C 
      \end{eqnarray}
is equal to  the area of the spherical surface subtended 
by the unit vector $ \hat{\bm d}={\bm d}/|{\bm d}|$ when integrating over the whole  2D Brillouin zone.  Here $C$ is the Chern number, which is equal to 1  in the topological-insulator phase $M>0$, but vanishes  in the conventional-insulator phase $M<0$, in full agreement with topological theorems.  This is illustrated in Fig.~\ref{figM}. 

After integration the single state current  (\ref{trCaur}) over the $\bm k$ space and  summation of the contributions of the two blocks in the Hamiltonian (\ref{ham4}) with opposite directions of spin,  only the topological term 
contributes to the total spin current $J^z_y=\sigma_s E$. Eventually the spin conductance is
\be
\sigma_s ={2 e  \bar s^z\over h} C.
    \ee{sc}
This derivation of the quantum spin conductance is well known \cite{BHZ}, but it is valid only for periodical boundary conditions. 

\section{Effect of reflective borders on the bulk spin current} \label{RBC}

\begin{figure}
\begin{center}
\begin{tabular}{c}
   \includegraphics[height=10cm]{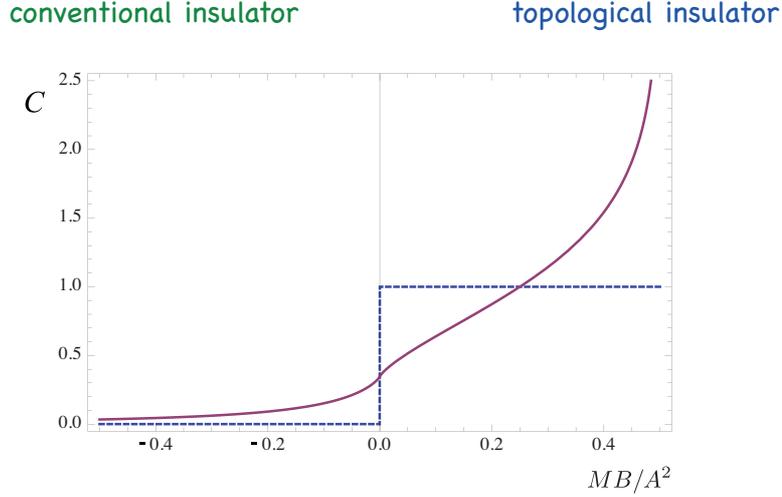} 
\end{tabular}
\end{center}
 \caption{ The parameter $C$ determining the spin conductance. For the periodic boundary condition $C$ reduces  to the Chern numbers $C=1$ and $C=0$  in the topological and the conventional phases respectively (dotted line). For the reflective boundary condition $C$ is continuous and does not vanish in the both phases (solid line).}
 \label{figCh}
 \end{figure}

Let us consider now fully reflective edges of the 2D topological insulator. If the spins of two bands are equal, i.e., 
$ \Delta s^z=0$ in the expression  \eq{trCaur} for the spin current, the spin current is proportional to the charge current. In the   BHZ model  spin-flips processes are absent [no off-diagonal blocks in the Hamiltonian \eq{ham4}],  the total number of electrons with given spin is conserved, and their current  normal to the reflective border should vanish. Correspondingly the spin current, which is a counterflow of electrons with opposite spins must also vanish. So the bulk spin  current is possible only due to the second term $\propto \Delta s^z$ in   \eq{trCaur}. A proper eigenfunction near the reflective border is a superposition of an incident and a reflected wave $a_1 \tilde{\bm \Psi}_\pm(k_x,k_y)+a_2 \tilde{\bm \Psi}_\pm(k_x,-k_y)$. The superposition must satisfy the condition 
$|a_1|^2 J_y(k_x,k_y)+|a_2|^2 J_y(k_x,-k_y)=0$ imposed by the charge conservation law. Here $J_y(k_x,k_y)$ is the charge current component normal to the border, which is determined by \eq{ChCur} for the wave function \eq{psiE}  taking into account the electric field. In the presence of the electric field $|a_1 |^2$ and $|a_2 |^2$ are not equal.  According to Eq.~(\ref{trCaur}) and  assuming that the electric field does not change  the average density ($|a_1 |^2+|a_2|^2=2$), the intensities are
\begin{eqnarray}
|a_1 |^2= \left(1-{ eE  \over 2 \hbar}{{\cal G}\over v_y}\right),~~|a_2 |^2= \left(1+{ eE  \over 2 \hbar}{{\cal G}\over v_y}\right).
                \end{eqnarray}
As a result, the term in the spin current proportional to the average spin $ \bar s^z$ vanishes but the term proportional to the spin difference $\Delta s^z$ still remains:
\begin{eqnarray}
J_y^z(k_x,k_y) =  \Delta s^z v_{0y}(|a_1 |^2-|a_2|^2)= \Delta s^z { eE  \over 2 \hbar}{\cal G}{v_{0y}\over v_y}.
   \end{eqnarray}
The total spin current in the whole band does not reduces to the Chern term and is determined by the integral, which does not vanish in a conventional insulator ($M<0$): \footnote{One cannot extend this analysis on $M>A^2/2B$ since in this case there are two real values of $k_y^2$  for the same energy (see Fig.~\ref{f1}b). So there are two reflected propagating waves,  and the spin current cannot be calculated without a detailed analysis of the wave function near the edge. }
\begin{eqnarray}
\int {\cal G}{v_{0y}\over v_y} d{\bm k}=2\pi \left\{\begin{array}{cc}  \left(-{A\over 2\sqrt{A^2-4MB}}\ln\frac{A^2-2MB+A\sqrt{A^2-4MB} }{A^2-2MB-A\sqrt{A^2-4MB}}
+\ln {A^2-2MB\over 2|M|B}\right) & \mbox{at}~M<{A^2\over 4B} \\  \left(-{A\over \sqrt{4MB-A^2}}\arctan \frac{A\sqrt{4MB-A^2}}{A^2-2M B}
+\ln {A^2-2MB\over 2|M|B}\right) &  \mbox{at}~{A^2\over 4B}<M<{A^2\over 2B}\end{array} \right. .
      \end{eqnarray}
The parameter $C$ for the periodic and reflective boundary conditions  is shown in Fig.~\ref{figCh}.
Thus, in contrast to the analysis based on plane-wave eigenstates satisfying the periodic boundary condition, the bulk spin current may appear both in the conventional and the topological insulator, and is not governed by the Chern number if the edge of the sample is fully reflective.

\section{Bulk spin current and spin accumulation} \label{spAc}

Observation of the calculated bulk spin conductance   in the experiment encounters the problem that  in an ideal topological insulator there is no electric field in the bulk since ballistic edge channels, which are robust against elastic scattering, shortcircuit  the bulk transforming a topological insulator into a ballistic metal. According to the Landauer--B\"uttiker theory of ballistic conductors  the whole voltage bias between electrodes drops  only in the contacts but not in the bulk (see Fig.~\ref{figTI}(b)). Correspondingly the bulk spin current linear in the electric field cannot appear and it is impossible to measure quantum spin conductance (i.e., to demonstrate the QSH effect in its original definition) in the strip geometry shown in Fig.~\ref{figTI}.  

The electric field parallel to the axis $x$ may appear only if there are magnetic impurities at the sample edge, which make carrier backscattering possible and lead to finite conductance  of edge channels. However in the absence of the spin conservation law the bulk current not necessarily leads to accumulation. It may result in an edge torque without accumulation, as in the case of equilibrium spin currents in the Rashba medium \cite{Adv}.  
As already mentioned in Introduction non-equilibrium spin accumulation is impossible also in insulators since the electric field does not pump energy into the material without longitudinal charge current. The BHZ model is a model of non-interacting electrons, and the band electrons in the bulk and the electrons in the edge states constitute two independent non-interacting ensembles. The bulk band electrons are in the insulating phase and cannot produce any accumulation of spin. Thus independently from
 whether the electric field and the related bulk spin current  do appear inside the sample or do not,  it cannot lead to spin accumulation at the edges. 
  Nevertheless,           it does not rule out spin accumulation in the edge states, which is inevitably appears due to a rather trivial reason: the voltage drop between electrodes generates  a ballistic electric current $J= e v_e(n_\rightarrow- n_\leftarrow)$ in edge channels,   which leads to inequality of the rightmover and leftmover densities $n_\rightarrow$ and $ n_\leftarrow$. Since  rightmovers and leftmovers have opposite spins this leads to  spin  accumulation (or polarization),
\begin{eqnarray}
S_z= {\bar s^z\over e v_e}J,
 \label{Macc}  \end{eqnarray}
totally independent from what is going on with  bulk spin currents. Only the average spin appears in this expression since according to \eq{edg}  the edge state is a superposition of two bands with equal amplitudes. 

One can rewrite \eq{Macc} for the edge spin polarization using the relation connecting  the edge channel current $J=(e^2/h)V$ with the voltage between electrodes via the charge quantum conductance $e^2/h$:
\begin{eqnarray}
S_z= {\bar s^ze \over h }{V\over v_e}.  
   \end{eqnarray}
where the half of the quantum spin conductance appears as a factor.   One may argue that since spin polarization contains the quantum of spin conductance its measurement would be a demonstration of the QSH effect. Such conclusion is at least debatable. The conductance quantum  appears in this expression because of charge conductance of the ballistic transport  without any connection with the actual bulk spin current, which is proportional to $\Delta s_z$ whereas the spin polarization is proportional to the average spin $\bar s^z$. The choice of the nomenclature is always  a matter of taste and convention and one may redefine the QSH effect as the effect of  spin polarization. But such a definition would be very far from the original concept of the SH effect \cite{DP}, which considered accumulation of spin brought to the edge by the bulk spin current linear in the bulk electric field.  For comparison, let us mention that the quantum charge conductance appears not only in the quantum Hall (now we mean not spin Hall!) effect, but also in other phenomena,  e.g. in the ballistic transport in 1D channels (the edge state is an example of such a channel). But this does not make these phenomena identical to the quantum Hall effect.

\section{Conclusion: observation of the genuine QSH effect} \label{concl}

\begin{figure}
\begin{center}
\begin{tabular}{c}
   \includegraphics[height=7cm]{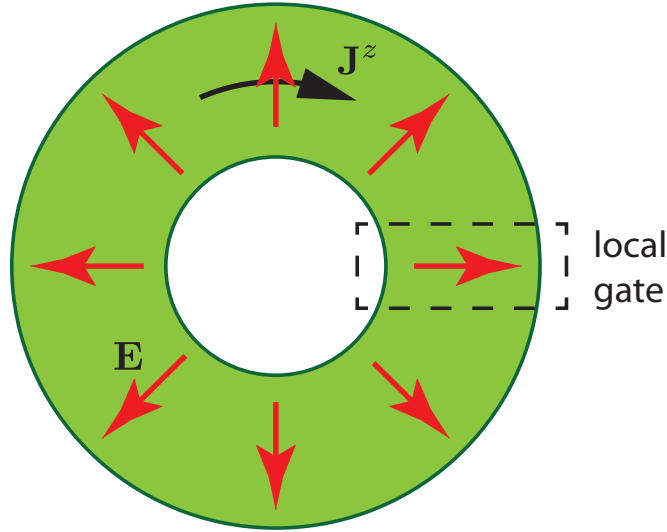} 
\end{tabular}
\end{center}
 \caption{ Tentative set up for observation of the QSH effect in the Corbino geometry. The radial electric field $\bm E$ induces the azimuthal spin current $\bm J^z$. Applying the voltage to the local gate (dash-line rectangle)  one can create a potential barrier transforming the periodic boundary condition into the reflective one.}
 \label{figCor}
 \end{figure}

The analysis presented above leads to the conclusion that the genuine QSH effect (bulk quantum spin conductance transverse to an external electric field) cannot be detected not only from transport measurements of charge conductance but even from observation of spin polarization at sample edges reported recently \cite{SpPol}. Indeed, the quantum spin conductance determined by the Chern number is not relevant for this observation and was not used for its interpretation.  The detection of the QSH effect
 is possible only if experimentalists are able to probe the bulk spin current in the geometry with the periodic boundary conditions when there are no edges parallel to the electric field and correspondingly no edge spin polarization . There is no doubt that it  is a serious challenge for experimentalists. 
Nevertheless such a geometry can be considered, at least in the theory. Figure~\ref{figCor} shows the Corbino geometry for a HgTe quantum well, where the transition to the topological phase is possible. Applying a radial electric field one creates an azimuthal  spin current. The Corbino geometry provides the periodic boundary condition, which allows to observe the quantum spin conductance determined by the Chern number. With help of a local gate (see the figure) one can create a potential barrier, which leads  to reflection of electrons propagating azimuthally and therefore can transform the boundary condition from  periodic  to purely reflective. This would make possible to experimentally check two predictions: quantum spin conductance for periodic boundary conditions, i.e., the genuine QSH effect,  and non-quantum spin conductance for reflective boundary condition, which may exist both in a conventional and a topological insulator.  An important feature of the Corbino geometry with the periodic boundary condition is that there are no ballistic helical edge channels, which are parallel to the electric field and shortcircuit the electrodes. So the electric field is present in the sample.

The most difficult part is detection of a bulk spin current in the absence of any spin polarization. 
 A possible method of bulk moment current detection is observation of an electric field generated by any moving magnetic moment  \cite{Adv,Zhang,Nag}. 
This is the ``inverse spin Hall effect", which has already been observed in a number of experiments 
\cite{VT,ISH,Ando,Tart}. But in the geometry of the QSH effect the electric field generated by the spin current  is exactly  parallel to the external electric field. So measurement of the spin current requires detection of a small correction to the dielectric permeability. The local gate can help to this detection, since it works as a valve, which regulates the spin current and therefore tunes the small correction to the dielectric permeability. 

One may also hope to find suitable optical methods of spin-current detection. Recently a second-order nonlinear optical effect of spin currents was revealed theoretically \cite{optT} and experimentally \cite{optic}. It was demonstrated that a pure spin current, even without spin polarization, is able to induce a second-harmonic generation. Concluding, the experimental detection of the Chern number, which determines the quantum spin conductance, seems elusive at the present moment, and some new ingenious set-ups should be looked for this goal.

\acknowledgments
I thank Laurens Molenkamp and Shou-Cheng Zhang for interesting discussions and giving their latest preprint \cite {SpPol} before its submittal to the cond-math archive. I also appreciate an  important  comment by Hui Zhao concerning the nonlinear optical effect of spin currents \cite{optT,optic}. 
The work was supported by the grant of the Israel Academy of Sciences and Humanities.

\bibliographystyle{spiebib}   

\end{document}